\documentclass[12pt]{iopart}
\usepackage{iopams}
\usepackage{setstack}
\usepackage{graphicx}
\usepackage{subfigure}
\usepackage{bm}
\begin{document}
\title[Properties of the one-dimensional Hubbard model]{
Properties of the one-dimensional Hubbard model: cellular dynamical mean-field description
}
\author{Ara Go and Gun Sang Jeon\footnote{
Author to whom any correspondence should be addressed.
}
}
\address{Department of Physics and Astronomy, Seoul National University, Seoul
151-747, Korea}
\ead{gsjeon@phya.snu.ac.kr}


\begin{abstract}
The one-dimensional half-filled Hubbard model is considered  at zero temperature
within the cellular dynamical mean-field theory (CDMFT). 
By the computation of 
the spectral gap and the energy density with various cluster and bath sizes
we examine the accuracy of the CDMFT in a systematic way, which proves the
accurate description of the one-dimensional systems by the CDMFT with 
small clusters.
We also calculate 
the spectral weights in a full range of the momentum for various interaction
strengths.
The results do not only account for the spin-charge separation, 
but they also 
reproduce all the features of the Bethe ansatz dispersions,
implying that
the CDMFT provides an excellent description of 
the spectral properties of low-dimensional interacting systems.

\end{abstract}

\submitto{\JPCM}


\section{Introduction}

Strong correlations arising from the electron-electron interactions have been one of the 
key issues to solve in the condensed matter physics.
The study on the strong correlations was initiated for the description of
interacting electrons in localized orbitals. 
It has been accelerated by
the discovery of interesting correlation properties such as high-temperature 
superconductivity or the colossal magnetic resistance. 

The main difficulty in the study of electron correlations is
that even the simplified theoretical models, which include only a small 
number of important degrees of freedom, 
are very challenging to solve.
The advent of the dynamical mean-field theory (DMFT)~\cite{Georges1996}
 has enabled a successful description of many important properties of strongly 
correlated systems in a unified frame.
Such a success was accomplished since the DMFT fully includes the local quantum
dynamics.  
On the other hand, the spatial correlations are treated in the mean-field level 
within the DMFT, and accordingly the DMFT is exact only in the limit of
infinite dimensions.

Intensive efforts to improve the accuracy of the DMFT in finite dimensions 
have been made in many directions.
Its cluster extensions~\cite{Maier2005} such as the cellular dynamical mean-field theory
(CDMFT)~\cite{Kotliar2001,Bolech2003}, 
the dynamical cluster approximation~\cite{Hettler2000}, 
and the variational cluster approach~\cite{Potthoff2003}
include short-range correlations inside the cluster with the mean-field
treatment of correlations between the clusters.
The effects of spatial correlations are taken into account step by step as the
cluster size is increased. 
The perturbative inclusion of the spatial correlations in a different direction
has been tried by the diagrammatic expansions such as the dual-fermion
method~\cite{Rubtsov2008} and the dynamical vertex approximation~\cite{Toschi2007}.
The fluctuations due to spatial correlations are most severe in one dimension,
leading to the expectation that 
one-dimensional interacting systems will provide strict tests
for the efficiency of the attempts to incorporate the effects of spatial correlations. 
The Hubbard model is particularly useful since its exact solution is known in
one dimension via the Bethe ansatz method~\cite{Lieb1968}.

The main purpose of our paper is twofold:
We first examine systematically the improvement in the accuracy of 
the CDMFT with the increase of cluster and bath sizes.
Secondly, we verify 
the performance of the CDMFT in describing
the spectral properties of low-dimensional systems. 
For that purpose 
we investigate the properties of the one-dimensional Hubbard model 
at zero temperature
within the CDMFT with an exact diagonalization method used as 
an impurity solver. 
The comparison of the CDMFT results with the exact ones obtained 
from Bethe ansatz solution shows that
the CDMFT gives accurate results for 
the spectral gap and the energy density even with very small clusters used.
Although for small local interactions odd clusters yield a metallic phase which is not
present in one dimension, such a region is found to shrink quickly with the
increase of the cluster size.
Spectral weights computed from the CDMFT give a qualitatively good 
description of the spin-charge separation which is a characteristic of 
the one-dimensional interacting system. 
Furthermore they show excellent agreement with Bethe ansatz dispersions
as well as the remarkable consistency in their intensity with the existing 
experimental and theoretical results, indicating the high accuracy of spectral
properties within the CDMFT even with small clusters.
Similar conclustions on the validity of the CDMFT in the one-dimensional 
half-filled Hubbard model were reached in a previous 
work~\cite{Bolech2003} based on the investigation of spectral gaps.
Here we emphasize that
our analysis is extended significantly to the examination of 
the energy density and the spectral weights as well as the spectral gaps.
Furthermore, the effects of the cluster size and the number of bath sites
on the physical quantities are also clarified by the systematic study.

This paper is organized as follows:
section~\ref{sec.model} gives a description of the one-dimensional Hubbard model
and the method employed in this work.
The results for the properties of 
the one-dimensional Hubbard model are presented in section~\ref{sec.results}.
We give a summary in section~\ref{sec.summary}.

\section{Model and Method} \label{sec.model}

The Hubbard model is described by the Hamiltonian
\begin{equation}
        H= -t\sum_{\langle i,j \rangle\sigma} (c^{\dagger}_{i\sigma} c^{}_{j\sigma}+ {\rm h.c.}) + U \sum_i n_{i\uparrow}n_{i\downarrow} \nonumber 
        - \mu \sum_{i\sigma} n_{i\sigma},
\end{equation}
where $c^{\dagger}_{i\sigma}(c_{i\sigma})$ creates (destroys) an electron
with spin $\sigma$ at site $i$ and the number operator is defined 
by $n_{i\sigma} \equiv c^{\dagger}_{i\sigma}c_{i\sigma}$. 
The hopping amplitude between nearest neighbors, the on-site Coulomb repulsion, and the chemical potential are denoted by $t$, $U$, and $\mu$, respectively.
We will consider the half-filled case $\mu=U/2$ and express all the energy 
and the length scales in units of the hopping amplitude $t$ and the lattice spacing $a$ 
throughout the paper.

We use the CDMFT to investigate the properties of the Hubbard model in one dimension.
Within the CDMFT, the lattice problem is reduced to the interacting electrons
on the cluster of $N_c$ sites hybridized with the noninteracting bath.
The effective action $S_{\rm eff}$ for the cluster degrees of freedom is
constructed by integrating out all the other degrees of freedom:
\begin{eqnarray} 
        S_{\rm eff} &=& \int^{\beta}_{0} d\tau d\tau^\prime
        \sum_{\mu\nu\sigma}c^{\dagger}_{\mu\sigma} (\tau) \mathcal{G}^{-1}_{\mu\nu\sigma}( \tau - \tau^\prime ) c^{}_{\nu\sigma}(\tau^\prime) \nonumber \\
        &&+ \int^{\beta}_{0} d \tau \sum^{N_c}_{\mu=1}U n_{\mu\uparrow}(\tau) n_{\mu\downarrow}(\tau),
\label{eq.action}
\end{eqnarray}
where $\mu, \nu = 1,2,\cdots,N_c$ are the indices of the sites inside the cluster and
a Gaussian (dynamical) Weiss field $\mathcal{G}^{-1}$
accounts for the effects of the other sites.
The CDMFT imposes the self-consistency relation
\begin{equation} \label{eq.SC}
        \hat{\mathcal{G}}^{-1} (i\omega_n)
- \hat{\Sigma}^c(i\omega_n)= \hat{G}_{{\rm loc}}^{-1}(i\omega_n)
\end{equation}
which determines the Gaussian dynamical Weiss field
$\hat{\mathcal{G}}^{-1}$.
Here $\hat{\Sigma}^c$ is the cluster self-energy which we obtain by solving the
effective action in Eq.~(\ref{eq.action})
and $\hat{G}_{\rm loc}$ is a local
Green function for the cluster given by
\begin{equation}
        \hat{G}_{{\rm loc}}(i\omega_n) =  \sum_{\tilde{k}}
        \left[(i\omega_n + \mu) \hat{1} - \hat{t}(\tilde{k}) -
                \hat{\Sigma}^c(i\omega_n)
        \right]^{-1} ,
\end{equation}
where the momentum $\tilde{k}$ runs over a reduced Brillouin zone of the superlattice composed of the clusters,
$\hat{t}(\tilde{k})$ is the Fourier transform of the hopping matrix, and
the circumflex over symbols,~$\hat{\phantom{a}}$ , represents an $N_c\times N_c$ matrix form.

In order to obtain the cluster self-energy
we use a Hamiltonian formalism. 
We introduce an impurity Hamiltonian
\begin{eqnarray} \nonumber
        H_{{\rm imp}} &=& 
\sum_{\mu\nu\sigma}E_{\mu\nu} c^{\dagger}_{\mu\sigma}c^{}_{\nu\sigma} 
+ U \sum_\mu n_{\mu\uparrow}n_{\mu\downarrow} 
\\
&&        + \sum_{\mu l\sigma} (V_{\mu l \sigma}  a^{\dagger}_{l\sigma}c^{}_{\mu\sigma} + V^*_{\mu l \sigma}  c^{\dagger}_{\mu\sigma}a^{}_{l\sigma}) 
 + 
\sum_{l\sigma} \epsilon^{}_{l\sigma}a^{\dagger}_{l\sigma} a^{}_{l\sigma} ,
\label{eq.impH}
\end{eqnarray}
where $E_{\mu\nu}$ contains the hopping matrix elements inside
the cluster and the chemical potential. 
The impurity Hamiltonian
$H_{\rm imp}$ contains additional degrees of freedom
($l=1,2,\cdots,N_b$) on the bath which have energies
$\epsilon_{l\sigma}$ and are coupled with the cluster sites via
$V_{\mu l \sigma}$. 
The Gaussian Weiss field corresponding to the impurity Hamiltonian in 
Eq.~(\ref{eq.impH}) is

\begin{equation}
        [\mathcal{G}^{-1} (i\omega_n;\{\epsilon_{l\sigma},V_{\mu l\sigma}\})]_{\mu\nu\sigma} = i\omega_n \delta_{\mu\nu}
        - E_{\mu\nu} - \sum_l \frac{V^*_{\mu l\sigma}V^{}_{\nu l\sigma}}{i\omega_n -
                \epsilon_{l\sigma}},
\end{equation}
where $\delta_{\mu\nu}$ is the Kronecker delta. The parameters
$\{\epsilon_{l\sigma}, V_{\mu l \sigma}\}$ are self-consistently
determined by the relation in Eq.~(\ref{eq.SC}).

The self-consistency loop starts from guessing the parameters
$\{\epsilon_{l}, V_{\mu l}\}$.
(The spin indices are omitted for a simpler notation.)
We first compute the cluster Green function $G_{\mu\nu}$
by solving the impurity Hamiltonian through
the exact diagonalization of an impurity Hamiltonian matrix.
We then obtain the cluster self-energy from the Dyson's equation,
\begin{equation}
        \hat{\Sigma}^c = 
\hat{\mathcal{G}}^{-1} - \hat{G}^{-1},
\end{equation}
which in turn yields
the local Green function $\hat{G}_{{\rm loc}}$ in one dimension via
\begin{equation}
        \hat{G}_{{\rm loc}}(i\omega_n) =  \int^{\pi/N_c}_{-\pi/N_c}
        \left[(i\omega_n + \mu) \hat{1} - \hat{t}(\tilde{k}) -
                \hat{\Sigma}^c(i\omega_n)\right]^{-1} \frac{d\tilde{k}}{2\pi/N_c}  .
\end{equation}
The resulting local Green function $\hat{G}_{{\rm loc}}(i\omega_n)$
and the cluster self-energy $\hat{\Sigma}^c(i\omega_n)$ give a
        new Gaussian Weiss field $\hat{\mathcal{G}}_{{\rm new}}(i\omega_n)$
\begin{equation}
        \hat{\mathcal{G}}_{{\rm new}}^{-1}(i\omega_n) =
        \hat{G}_{{\rm loc}}^{-1}(i\omega_n) +
        \hat{\Sigma}^c(i\omega_n).
\end{equation}
We extract new parameters $\{\epsilon_{l}, V_{\mu l}\}$ by minimizing
the distance function $\chi^2$
\begin{eqnarray} \nonumber
    \chi^2 &\equiv& \frac{1}{N_{\rm max}+1} \sum^{N_{\rm max}}_{n=0}\sum^{N_c}_{\mu\nu}
        \left| [\mathcal{G}_{{\rm new}}^{-1}(i\omega_n)]_{\mu\nu}
			\right.
\\ && \left.
        - [\mathcal{G}^{-1}(i\omega_n;\{\epsilon_{l},V_{\mu l}\})]_{\mu\nu} \right|^2.
\end{eqnarray}
The distance function is evaluated on Matsubara frequencies
$\omega_n = (2n+1)/\beta$ with $n=0,1,\cdots,N_{\rm max}$
, $\beta=200$, and $N_{\rm max}=800$, and
a conjugate gradient algorithm is used in the minimization.
The self-consistency loop is repeated until the convergence is achieved.

\section{Results} \label{sec.results}
\begin{figure}[t]
    \centering \subfigure[]{\includegraphics[angle=-90,width=0.45\textwidth]{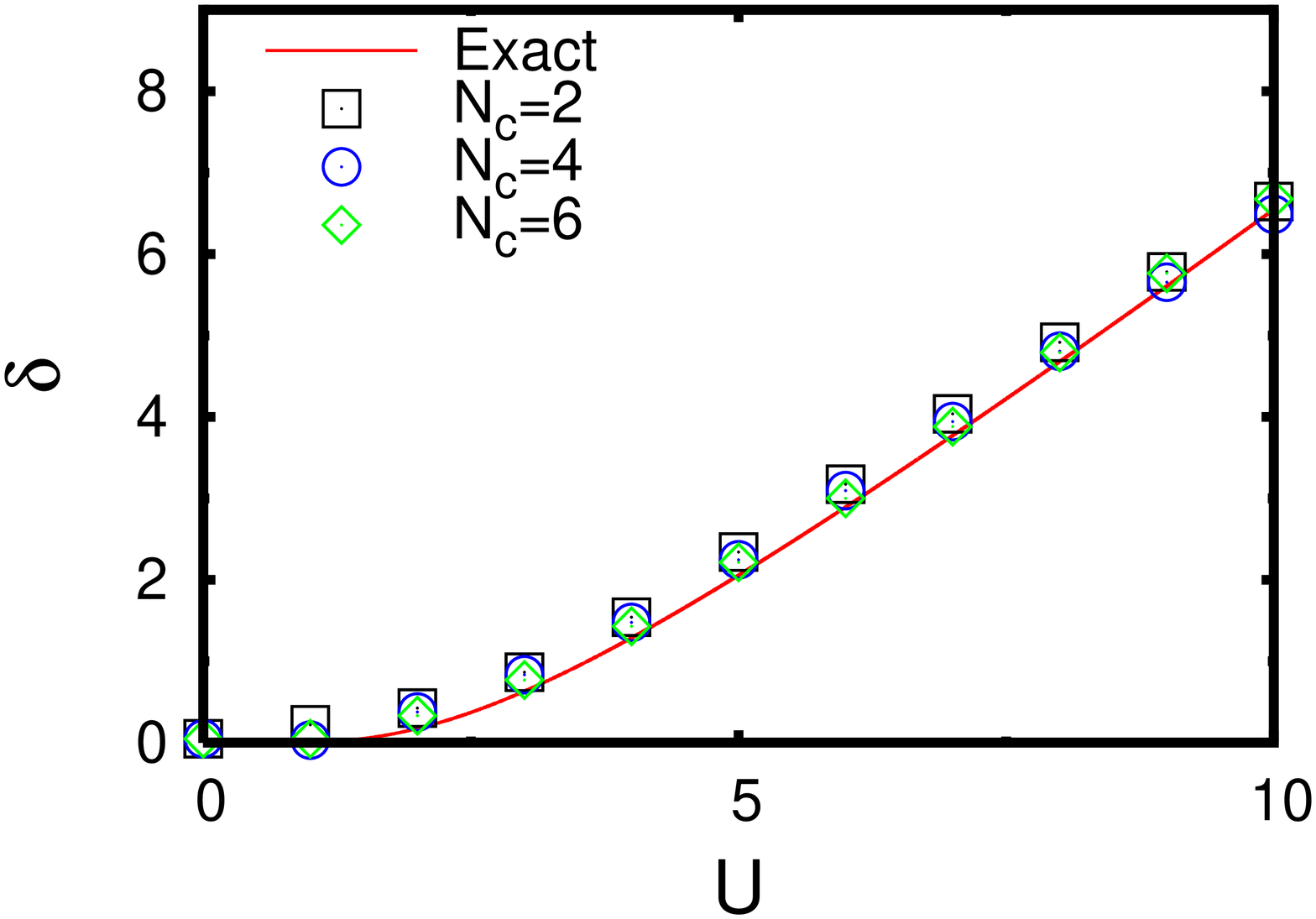}}
    \centering \subfigure[]{\includegraphics[angle=-90,width=0.45\textwidth]{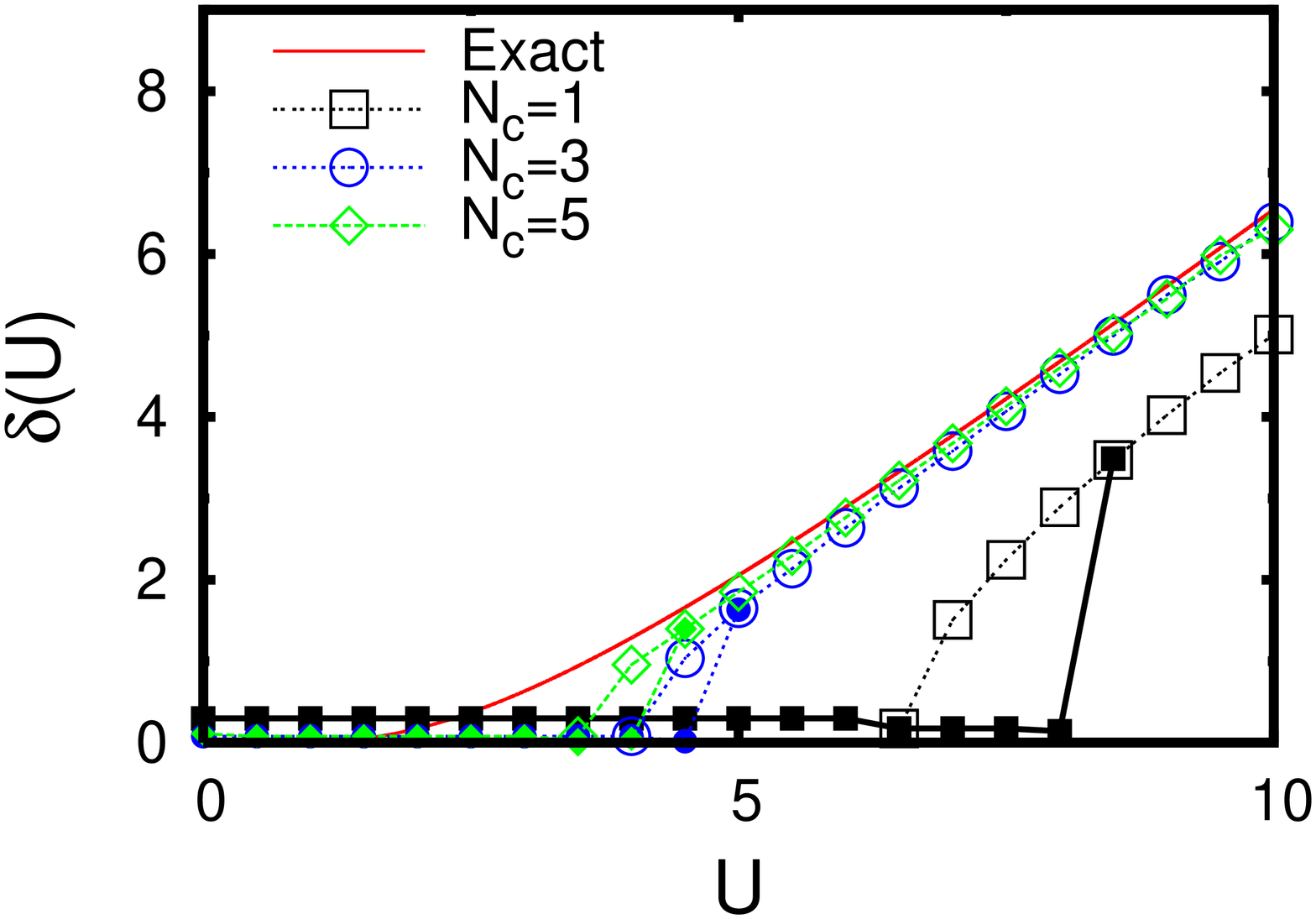}}
    \caption{\label{fig.gapNc} 
Spectral gap $\delta$ as a function of $U$ for (a) even $N_c$;
(b) odd $N_c$.
The gaps from the CDMFT calculation are denoted by symbols while the solid 
line represents an exact gap from the Bethe ansatz solution.
The empty symbols represent the spectral gaps obtained as $U$ is decreased 
while those obtained with the increase of $U$ are denoted by the 
filled symbols.
The number of bath sites $N_b$ is fixed to 6 for even $N_c$ and to 5 
for odd $N_c$. 
}
\end{figure}

We first calculate the spectral gap $\delta$ as a function of the interaction 
strength $U$ to demonstrate the accuracy of the CDMFT.
The gap is measured by the difference between the energy of 
the lowest electron peak and that of the highest hole peak in 
the single-particle density of states. 
We compare the results with an exact result given in a compact form
\begin{equation}
    \delta(U) = \frac{16t^2}{U} \int^\infty_1 \frac{\sqrt{y^2-1}}{\sinh(2\pi ty/U)} dy ,
    \label{eq.gap}
\end{equation}
which is obtained from the Bethe ansatz solution\cite{Ovchinnikov1970}. 
The CDMFT results of the spectral gap in the 
one-dimensional Hubbard model are available in an earlier work~\cite{Bolech2003}. 
In contrast to the previous work
we present the results for various cluster sizes $N_c$ and 
numbers of bath sites $N_b$ which reveals clearly the roles of 
$N_c$ and $N_b$. 

Figure~\ref{fig.gapNc} shows the spectral gap for various cluster size $N_c$.
For the clusters with an even number $N_c$ of sites, the CDMFT yields 
a spectral gap very close to an exact value, as shown in 
Fig.~\ref{fig.gapNc}(a).
The two-site cluster calculation already exhibits a qualitatively 
correct behavior of a spectral gap as a function of $U$, such as 
the appearance of a finite gap for every finite $U$.
Indeed the deviation of $N_c=2$ results from the exact values 
is quite small; such deviation is further reduced by an increase of $N_c$.
Unlike the earlier work~\cite{Bolech2003}
we have measured the spectral gap in the limit that the broadening 
of the peaks in the single-particle density of states vanishes. 
This limit discloses maximal numerical errors from the finite number of
cluster or bath sites. 
We can see that the spectral gap within the CDMFT is in excellent 
agreement with an exact one in such a worst limit.

\begin{figure}[t]
    \centering \subfigure[]{\includegraphics[angle=-90,width=0.45\textwidth]{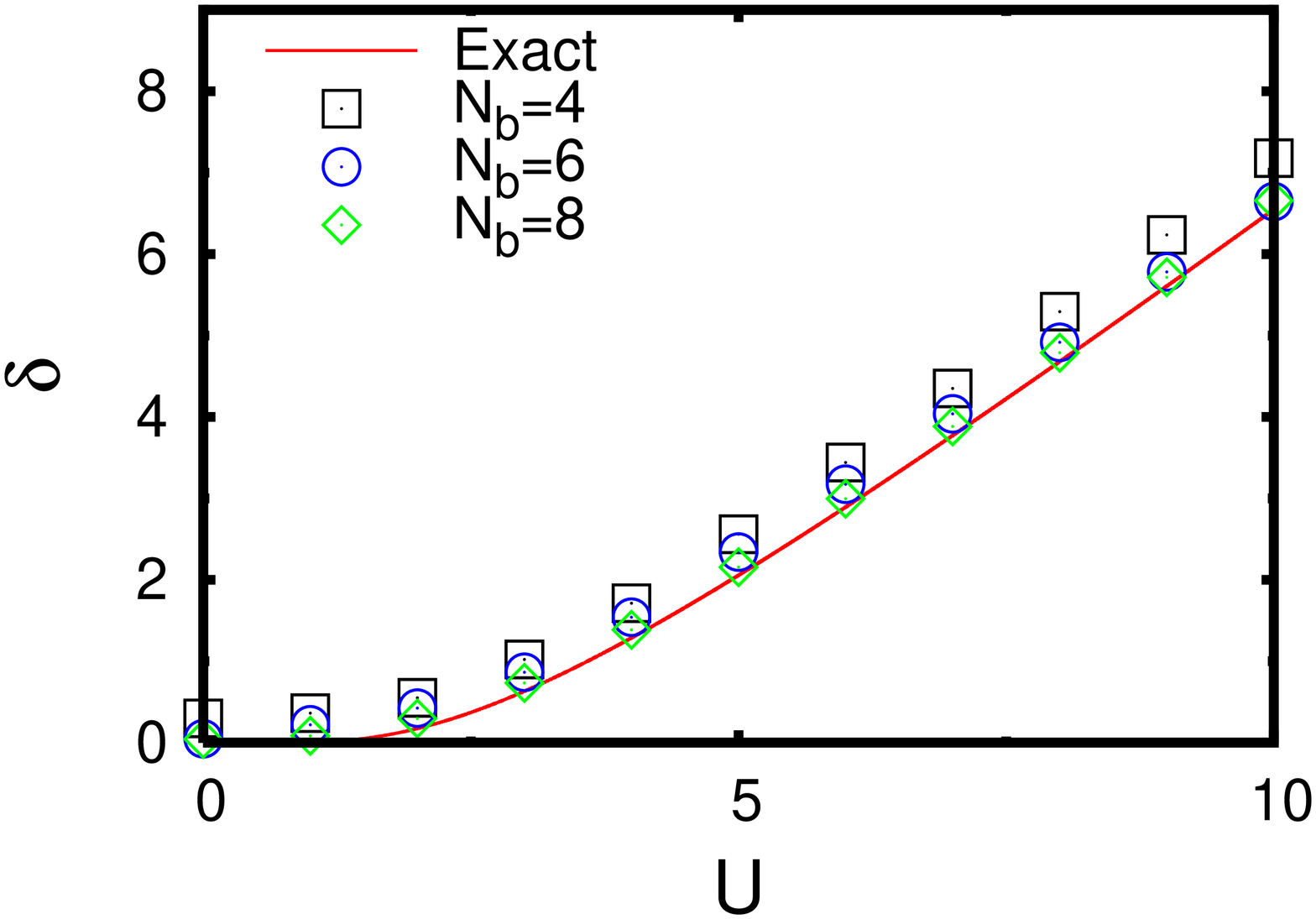}}
    \centering \subfigure[]{\includegraphics[angle=-90,width=0.45\textwidth]{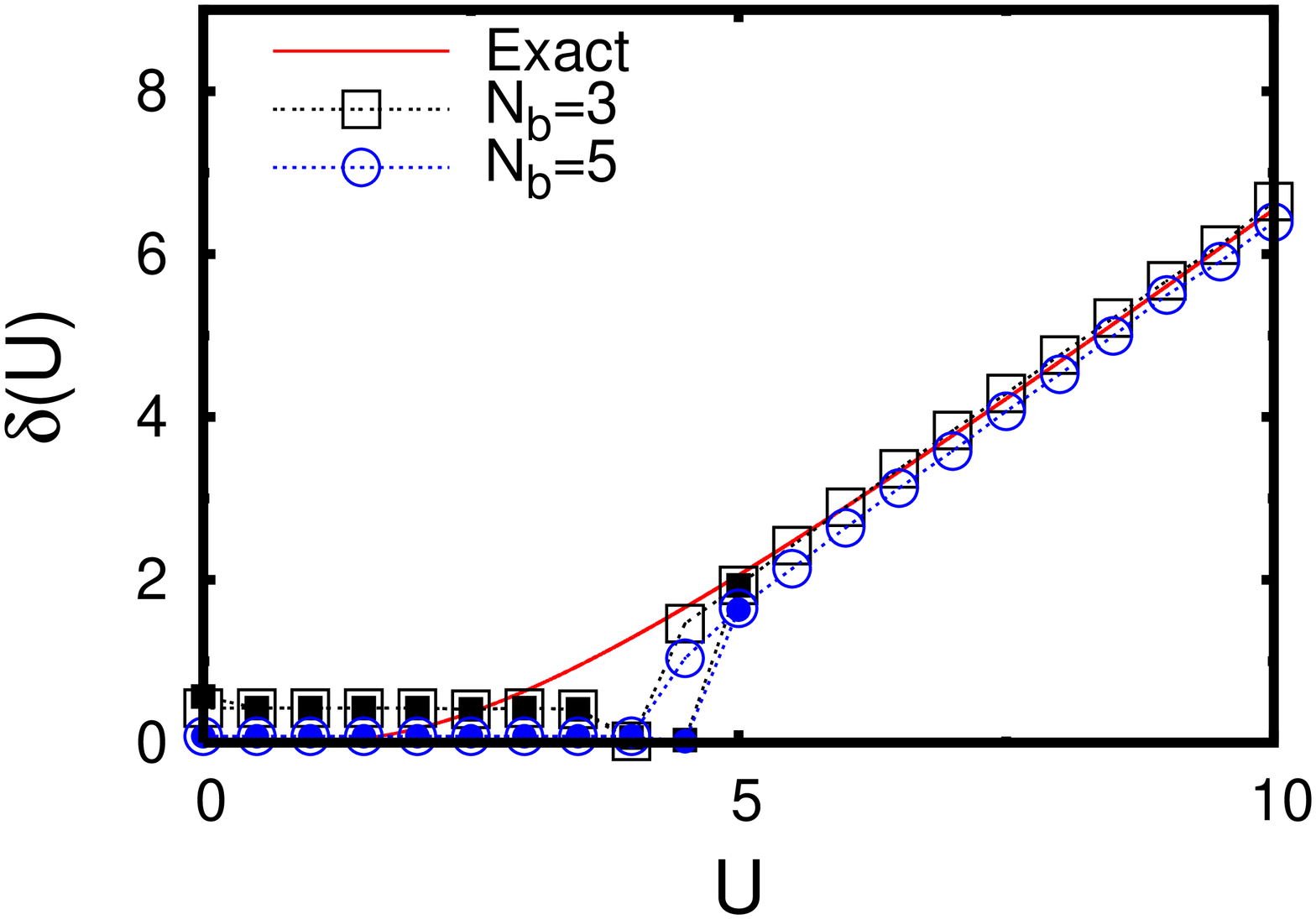}}
    \caption{\label{fig.gapNb} 
Spectral gap $\delta$ as a function of $U$ for various $N_b$ with (a) $N_c=2$;
(b) $N_c=3$.
The empty symbols represent the spectral gaps obtained as $U$ is decreased 
while those obtained with the increase of $U$ are denoted by the 
filled symbols.
The solid line represents an exact gap from the Bethe ansatz solution.
}
\end{figure}

\begin{figure}[t]
    \centering \subfigure[]{\includegraphics[angle=-90,width=0.45\textwidth]{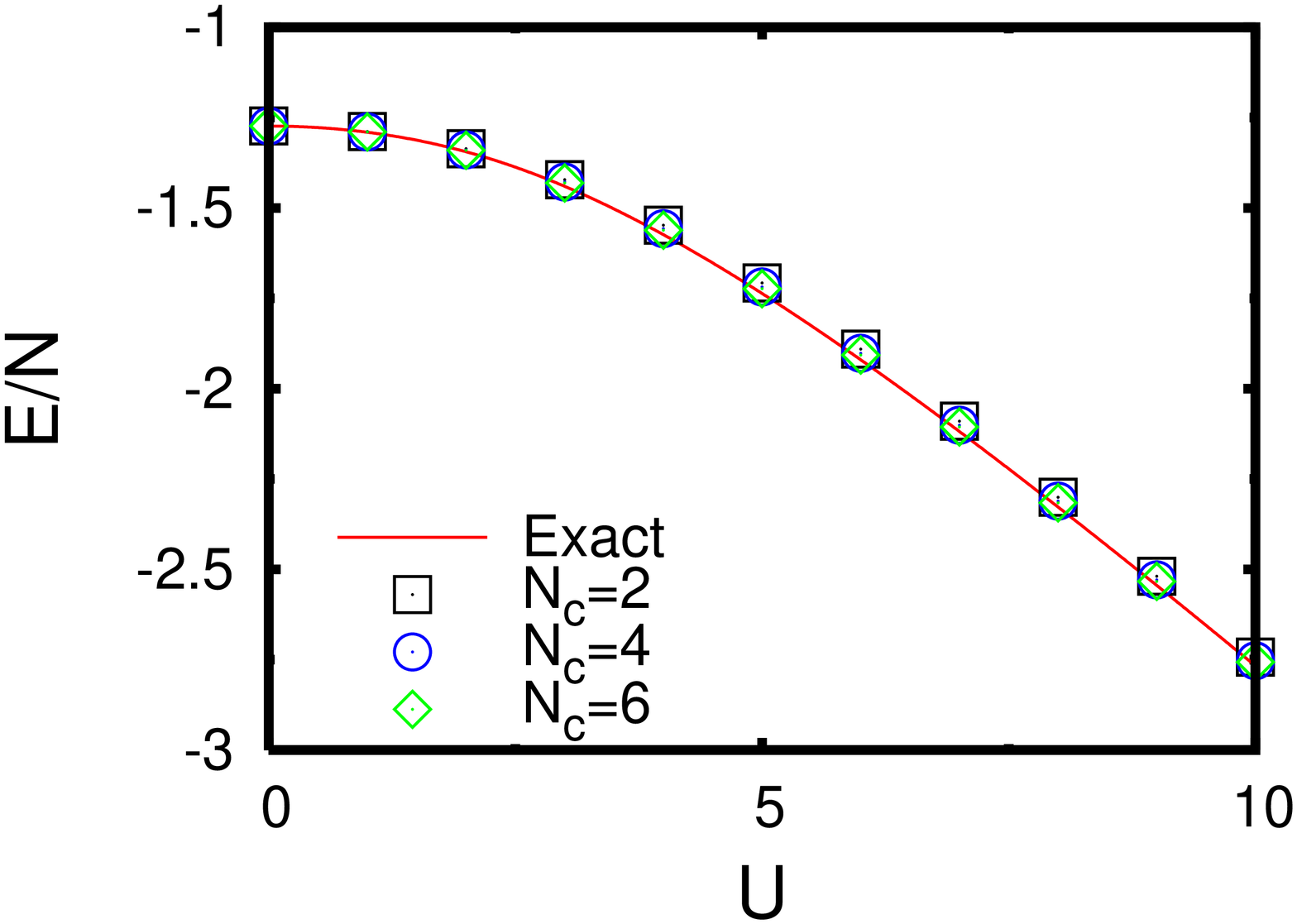}}
    \centering \subfigure[]{\includegraphics[angle=-90,width=0.45\textwidth]{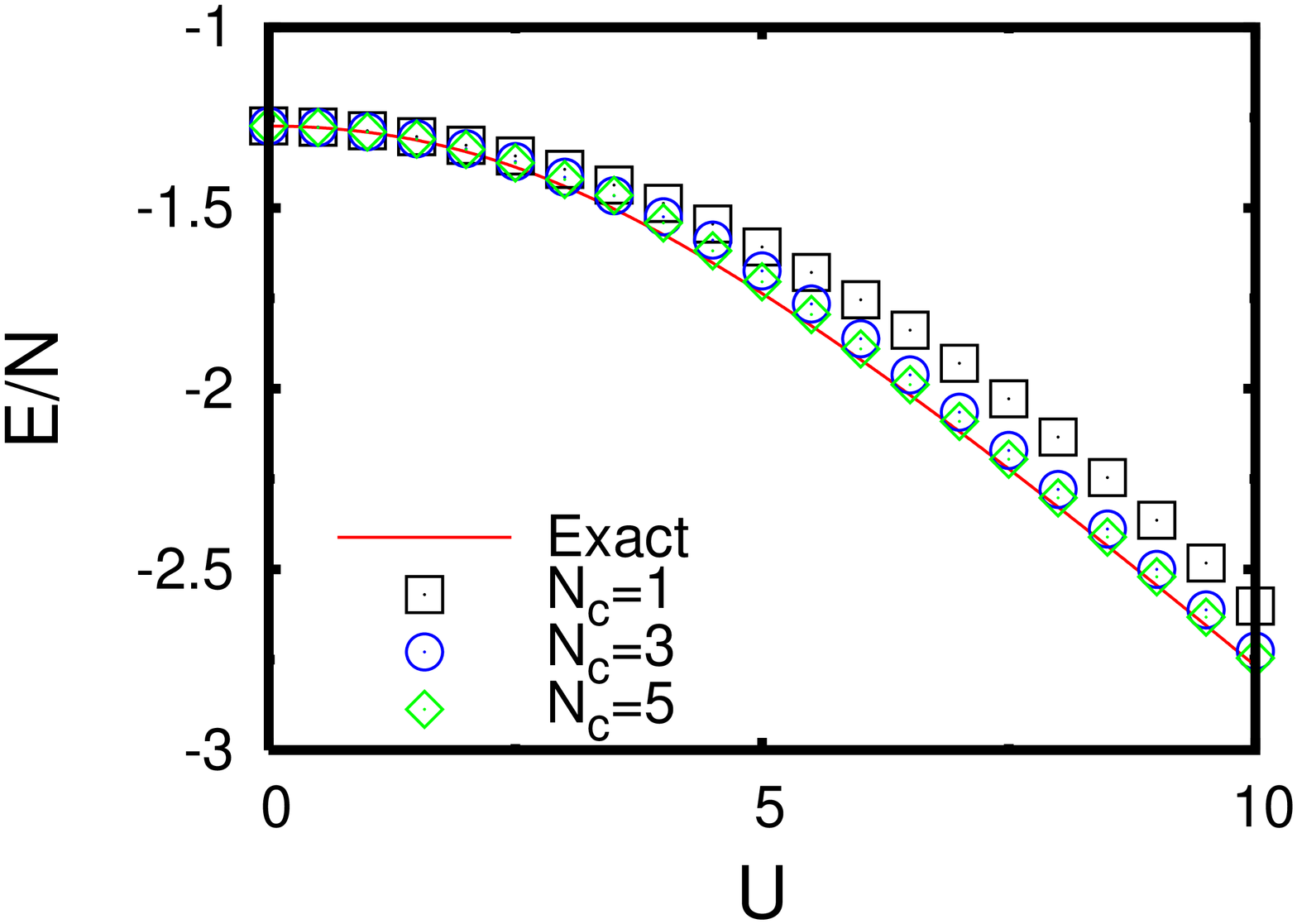}}
    \caption{\label{fig.energy}
    Ground state energy density $E/N$ as a function of $U$ for 
(a) even clusters with $N_b=6$, 
(b) odd clusters with $N_b=5$
within the CDMFT calculation.
The energy density with odd $N_c$ also shows a hysteresis in the coexistence 
region as the spectral gap does. 
Since the size of the hysteresis in the energy density is smaller than the 
symbol size in every system studied, we plot only the energy density obtained when $U$ is increased.
The exact energy density from the Bethe ansatz is represented by the solid
line for comparison.
    }
\end{figure}

Contrary to the case of even clusters,  the agreement
of spectral gaps from odd clusters ($N_c=1,3,5)$  with exact results 
is not good,
as shown in Fig.~\ref{fig.gapNc}(b).
They show a metallic phase over a finite region of small $U$
before entering a Mott phase through a coexistence region,
which is rather similar to the behavior in infinite dimensions.
Although odd clusters exhibit poorer agreement with exact results 
than even clusters, the increase of $N_c$ also causes a rapid reduction 
in the critical interaction strength together with the shrink of a coexistence 
region.
Indeed the cluster with $N_c=3$ reproduces essentially an exact gap  for
$U\gtrsim 6$.

We also plot spectral gaps for two- and three-site clusters with various 
numbers of bath sites.
As can be seen in Fig.~\ref{fig.gapNb},
the spectral gap computed within the CDMFT is also dependent on the number of
bath sites $N_b$.
This arises because the Gaussian Weiss field is approximated by a finite 
number of bath sites.
Accordingly,
the errors in the spectral gap decrease monotonically with $N_b$, which is
apparently faster for larger $U$.

\begin{figure}[t]
    \centering {\includegraphics[angle=-90,width=0.45\textwidth]{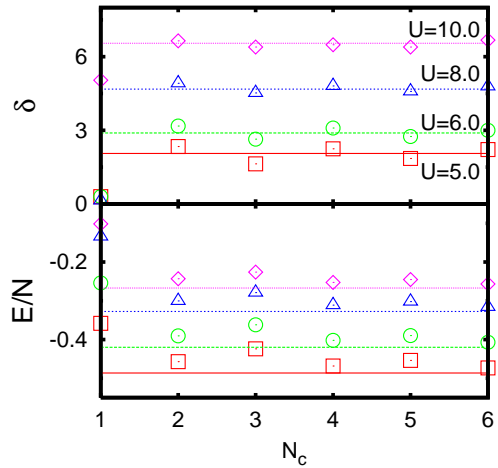}}
    \caption{\label{fig.ncvs} 
Spectral gap $\delta$ and ground state energy density $E/N$
as a function of the cluster size $N_c$ for various values of $U$.
    The results for $U=5, 6, 8, 10$ are denoted by the red squares, 
green circles, blue triangles, and pink diamonds, respectively.
    The lines represent exact values for the corresponding $U$ with the same
color.
    We used $N_b=5$ for odd $N_c$ and $N_b=6$ for even $N_c$.}
\end{figure}

In addition to the spectral gap we also examine the ground state energy 
density in order to check whether such accurateness of the CDMFT
shown in the study of spectral gaps persists in other physical quantities.
The ground state energy density is calculated as
\begin{eqnarray} \nonumber
\frac{E}{N} &= 
\frac{1}{\beta} \sum_{n} \sum_{\tilde{k}} \left[
   \frac{2}{N_c}{\rm Tr} 
\right.
&
\left\{
\left(
\hat{H}_0(\tilde{k})
+ \frac{1}{2} \hat{\Sigma}^c(i\omega_n)
\right)
\right.
\\
&&
\left. \left.
\phantom{\frac{2}{N}}
\times
[ (i\omega_n + \mu)\hat{1} - \hat{t}(\tilde{k}) -
\hat{\Sigma}^c(i\omega_n)]^{-1}
\right\}
\right]
    \label{eq.cenergy}
\end{eqnarray}
where $H_0$ includes kinetic and chemical potential terms in the
Hamiltonian
and the factor 2 comes from the spin degeneracy.
The summation over Matsubara frequency is transformed to an integral 
in the zero-temperature limit.
In Fig.~\ref{fig.energy} the energy densities within the CDMFT for various 
sizes of clusters are compared with an exact energy density. 
The exact energy density can also be calculated from the Bethe ansatz 
solution~\cite{Lieb1968} given by 
\begin{equation}
    \frac{E}{N} =
    -4 \int^\infty_0 \frac{J_0(\omega) J_1(\omega) d\omega}{\omega [1+ \exp(\frac{1}{2}\omega U)]}-\frac{U}{4},
    \label{eq.energy}
\end{equation}
where $J_n(x)$ are the $n$th-order Bessel functions of the first kind.
Except for the single-site calculation, all the calculations essentially 
reproduce the exact energy density.
Figure~{\ref{fig.ncvs} displays the explicit dependence of the spectral gap 
and the ground state energy density on the cluster size $N_c$.
Both quantities show the even-odd oscillations with the increase of
$N_c$, and 
gradually approach the exact value in the whole range of $U$.

\begin{figure}
    \centering {\includegraphics[angle=-90,width=0.45\textwidth]{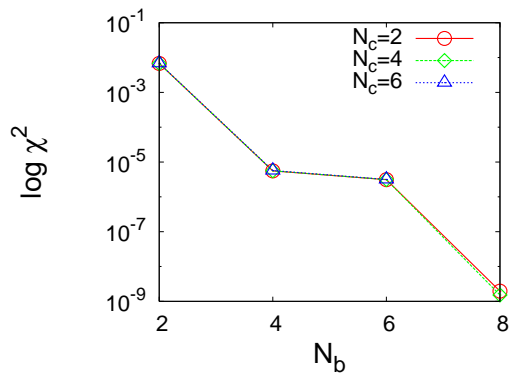}}
    \caption{\label{fig.chi}
    Distance function $\chi^2$ of the converged solutions for $U=8$ and various values of $N_c$ as a function of $N_b$.
    }
\end{figure}

The distance function $\chi^2$ for even clusters is 
plotted on a logarithmic scale 
as a function of $N_b$ in Fig.~{\ref{fig.chi}}.
Since the infinite bath is projected into the space spanned by 
$N_b$ sites in the exact diagonalization method,
larger $N_b$ is expected to give smaller distance function, 
implying better convergence.
In general the distance function $\chi^2$ decreases exponentially with increasing $N_b$.
A small bump at $N_b=6$ indicates that 
the baths of 4 and 8 sites are more favored than that of 6 sites in 
the calculation of the CDMFT.
This is because it is difficult for 6-site bath to reflect the inversion 
symmetry and the particle-hole symmetry of the system.
The effect of the cluster size $N_c$ on $\chi^2$ is negligible
because the limitation of the accuracy in the distance function comes mainly 
from the use of a finite number of bath sites.

\begin{figure}[t]
    \centering {\includegraphics[angle=-90,width=0.45\textwidth]{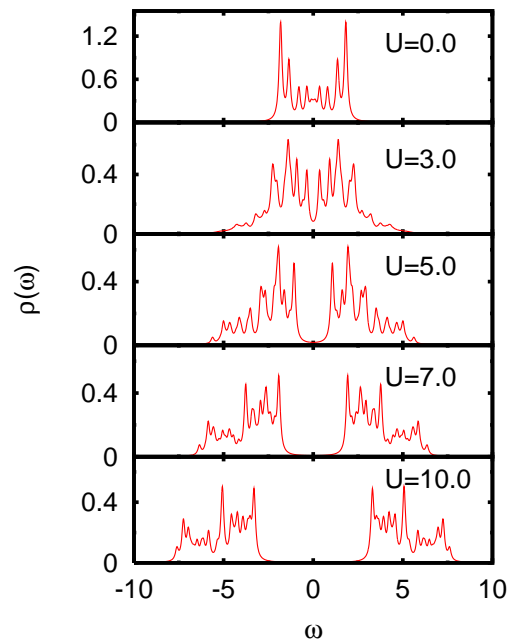}}
    \caption{\label{fig.dos}
Local density of states $\rho(\omega)$ for various values of 
$U$ for $N_c=4$ and $N_b=8$.
    We used a broadening factor $\varepsilon=0.1$.}
\end{figure}

Within the CDMFT we next calculate
the local density of states 
\begin{equation}
    \rho(\omega) = -\frac{1}{\pi}
 {\rm Im}[{\rm Tr}\hat{G}(\omega+i\varepsilon)],
\end{equation}
where $\varepsilon$ is a broadening factor.
We plot the local density of states for various $U$ in Fig.~{\ref{fig.dos}}.
The single band for $U=0$ is split to two bands, the upper and the lower
Hubbard bands, with the local interaction $U$ turned on.
The two bands are located around $\pm U/2$.
The Hubbard gap between the bands increases with the increase of $U$
and becomes of the order of $U$ in the strong interaction regime.
It is interesting to note that
the abrupt decrease of the density of states around the center of each Hubbard 
band separates two regions of the band, one with large density of state near
the gap and the other with small density of states,
which is more prominent for large $U$.
The analysis of the spectral weights presented below
provides an explanation for such a structure.

\begin{figure}[t]
    \centering \subfigure[]{\includegraphics[angle=-90,width=0.45\textwidth]{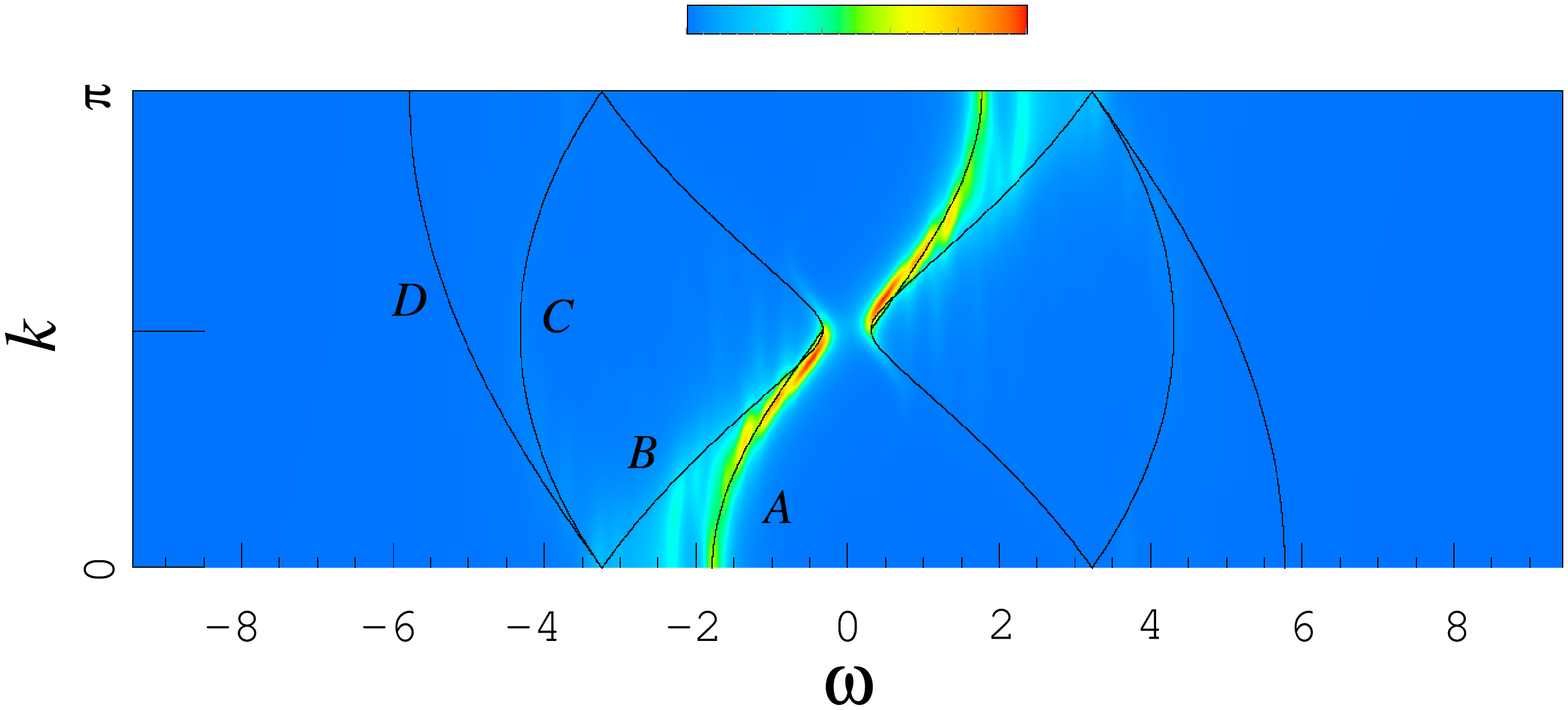}}\\
    \centering \subfigure[]{\includegraphics[angle=-90,width=0.45\textwidth]{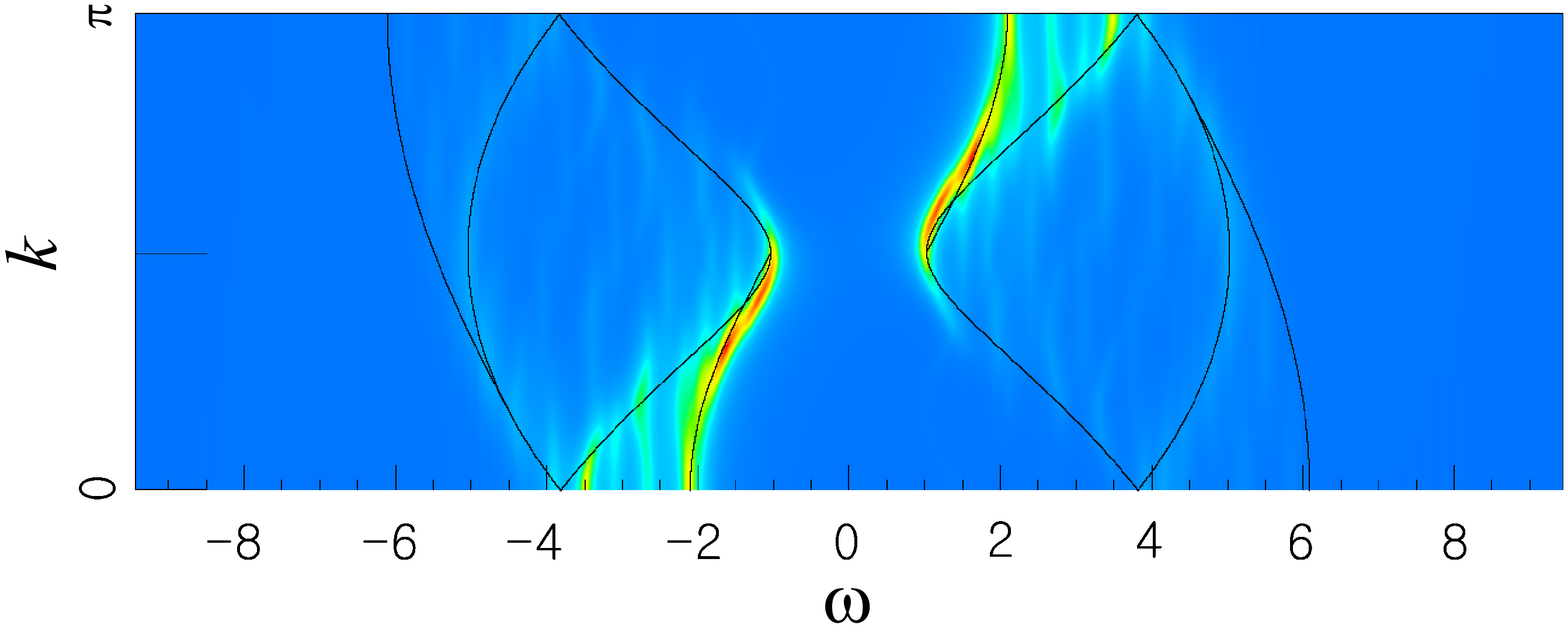}}\\
    \centering \subfigure[]{\includegraphics[angle=-90,width=0.45\textwidth]{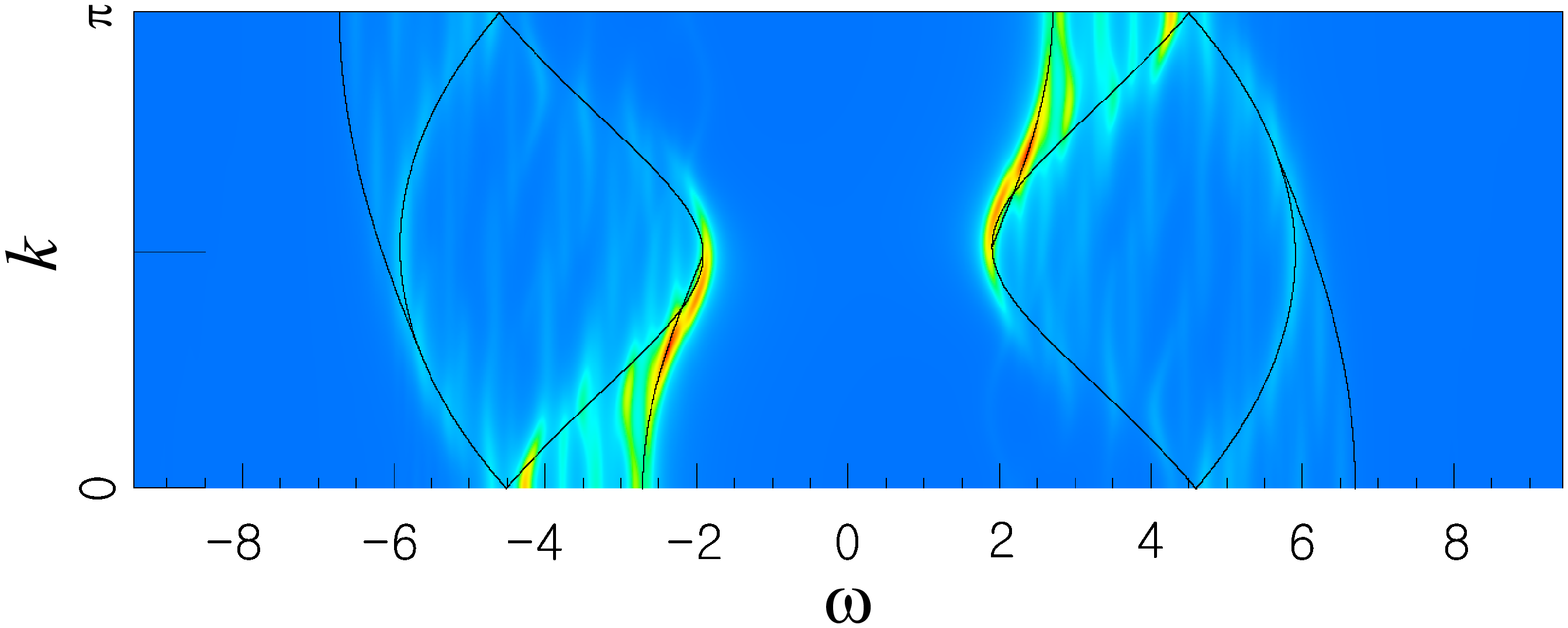}}
    \caption{\label{fig.sw}
The colored plot of spectral weight $A(k,\omega)$ for (a) $U=3.0$, (b) $U=5.0$
and (c) $U=7.0$.
	 We used the cluster of size $N_c=6$ with the bath of $N_b=6$ sites 
    and computed $A(k,\omega)$ with a broadening factor $\varepsilon=0.1$.
    We normalized each weight by its maximal value.
    Some important dispersions and boundaries of the spinon-holon continuum,
    which are obtained from the Bethe ansatz solution, are also represented by solid lines. 
(see the text for details).
    }
\end{figure}

The CDMFT allows us to compute
the spectral weight $A(k,\omega)$ as
\begin{equation}
    A( k, \omega) = - \frac{1}{\pi} {\rm Im} G_{\rm latt}(k,\omega+i\varepsilon)
    \label{eq.sw}
\end{equation}
with the lattice Green function
\begin{equation}
   G_{\rm latt}( k, \omega) = \frac{1}{N_c} \sum_{\mu\nu} e ^{i k (x_\mu - x_\nu ) }
       \bigg[ (\omega +\mu)\hat{1} - \hat{t}( k ) -
\hat{\Sigma}^c(\omega) \bigg]^{-1}_{\mu\nu}.
\end{equation}
Here $k$ is a vector in the original Brillouin zone and $x_\mu$ is the position of site $\mu$ in the cluster.
The spectral weight shown in Fig.~{\ref{fig.sw}}
gives an accurate description of
the spin-charge separation which is a peculiar feature of the
one-dimensional half-filled Hubbard model. 
When a hole or an electron is added to a one-dimensional interacting system, 
it is split into two collective modes,
 a spinon which is charge-neutral with spin 1/2 and
and a holon/antiholon which is spinless with a hole/electron charge.
The Bethe ansatz method provides only the dispersions of such elementary excitations
while their intensity could be calculated only in some special
limits~\cite{Sorella1992,Penc1996}.
Only recently has the reliable computation of the intensity been performed 
by dynamical density matrix renormalization
group~\cite{Benthien2004, Benthien2007}  and the cluster perturbation theory~\cite{Senechal2000}.
In Fig.~\ref{fig.sw} we marked some important branches: 
a spinon branch by $A$, a holon branch ($0<k<\pi/2$) and 
a secondary holon branch ($\pi/2<k<\pi$) by $B$,
the continuation of the holon branch from the negative-$k$ side by $C$,
and the lower boundary of the spinon-holon continuum by $D$.

The spectral weights computed for a 6-site cluster within the CDMFT 
exhibit an evolution with the variation of $U$ which is quite consistent
with Bethe ansatz dispersions.
We can see clearly that spinon ($A$) and holon ($B$) branches show up at the locations
predicted by the Bethe ansatz. 
Most weights are concentrated between the two branches, as has been observed
in the experiments~\cite{Kim2006, Koitzsch2006}.
As $U$ is increased, we observe some weights between the branches $B$ and $C$. 
Although some weights are found outside the branch $C$, they do not exceed the
boundary imposed by the branch $D$.
The fact that 
most spectral weights are located between the branches $A$ and $B$ explains 
the sudden jump in the density of states of the Hubbard bands which is 
mentioned above.
For $U=7$, for example,
the large density of states in the region $-4\lesssim \omega \lesssim -2$ 
has the main contribution from the region between the spinon branch $A$ and 
the holon branch $B$ while
the spectral weights outside the holon branch $B$ form a region of small 
density of states up to the lower boundary of the spinon-holon continuum 
( $\omega\approx-7$ ).

\begin{figure}[t]
    \centering \subfigure[]{\includegraphics[angle=0,width=0.55\textwidth]{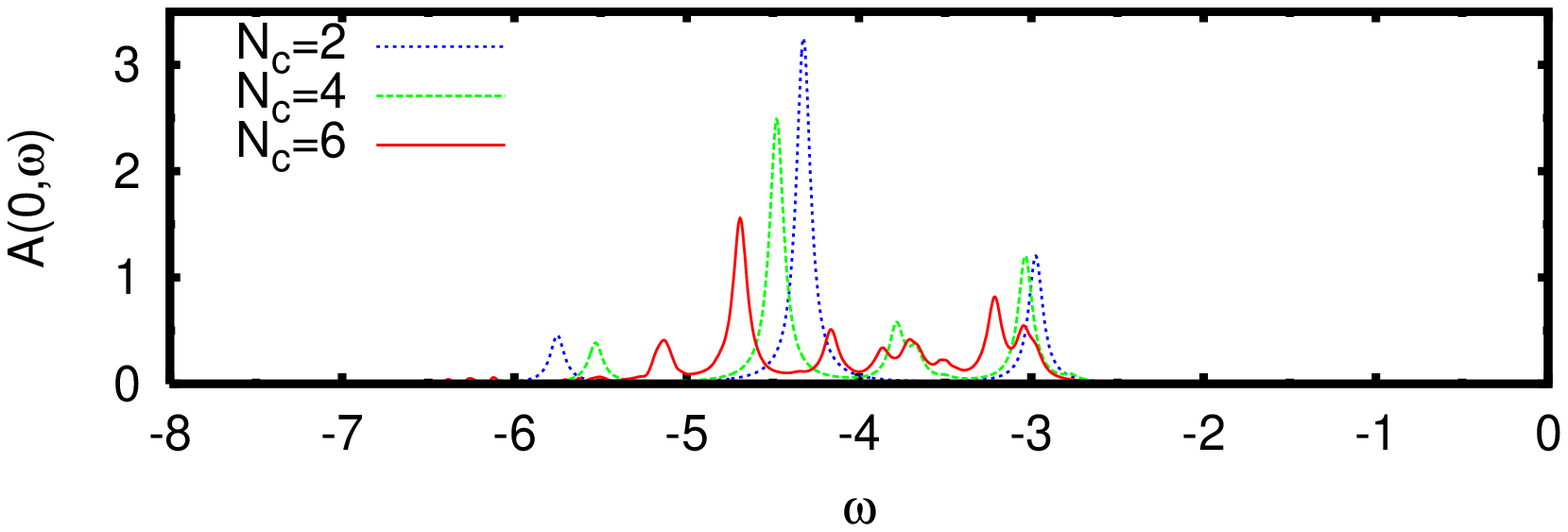}}\\
    \centering \subfigure[]{\includegraphics[angle=0,width=0.55\textwidth]{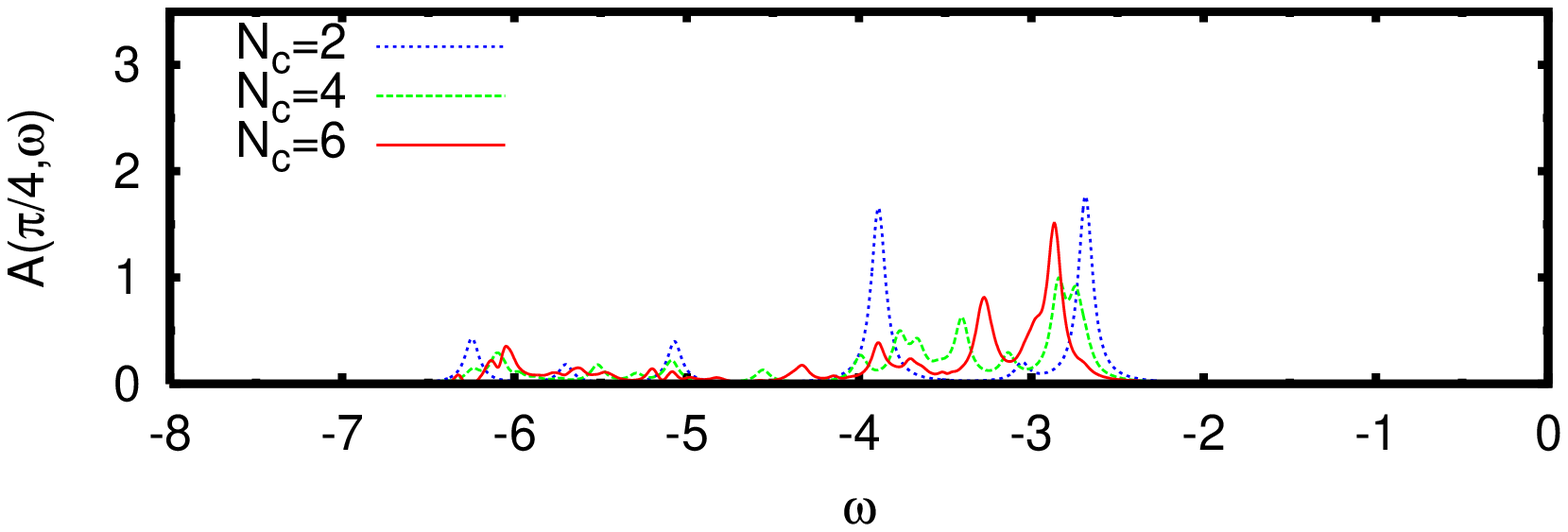}}\\
    \centering \subfigure[]{\includegraphics[angle=0,width=0.55\textwidth]{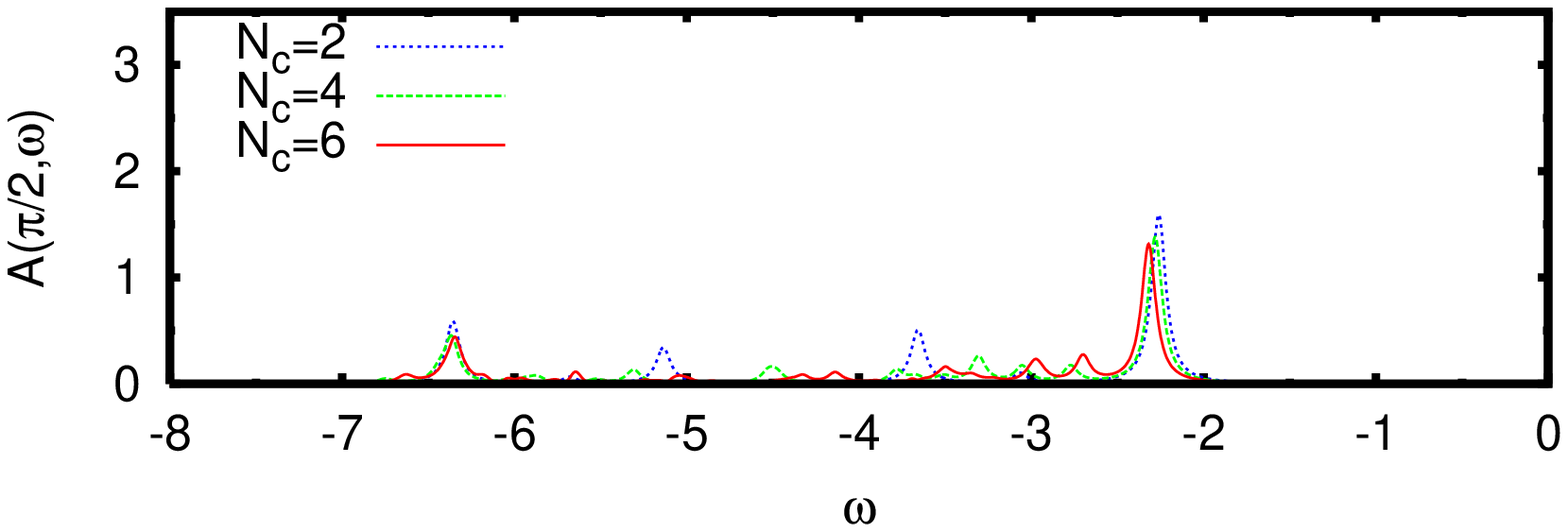}}\\
    \centering \subfigure[]{\includegraphics[angle=0,width=0.55\textwidth]{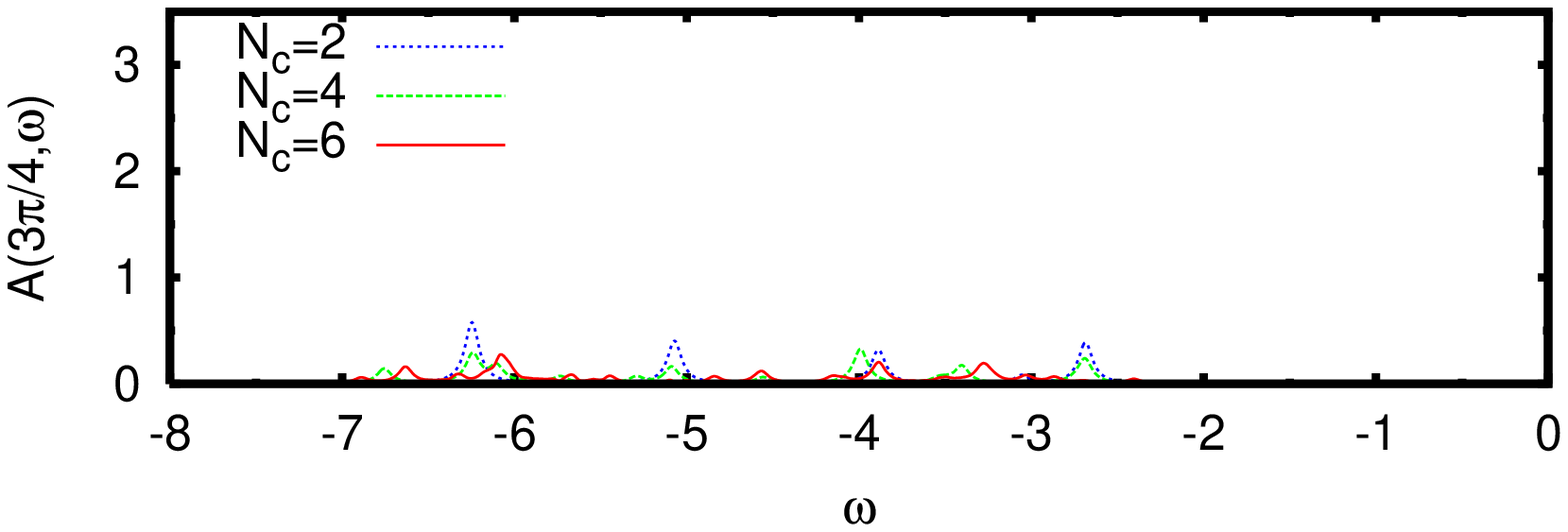}}\\
    \centering \subfigure[]{\includegraphics[angle=0,width=0.55\textwidth]{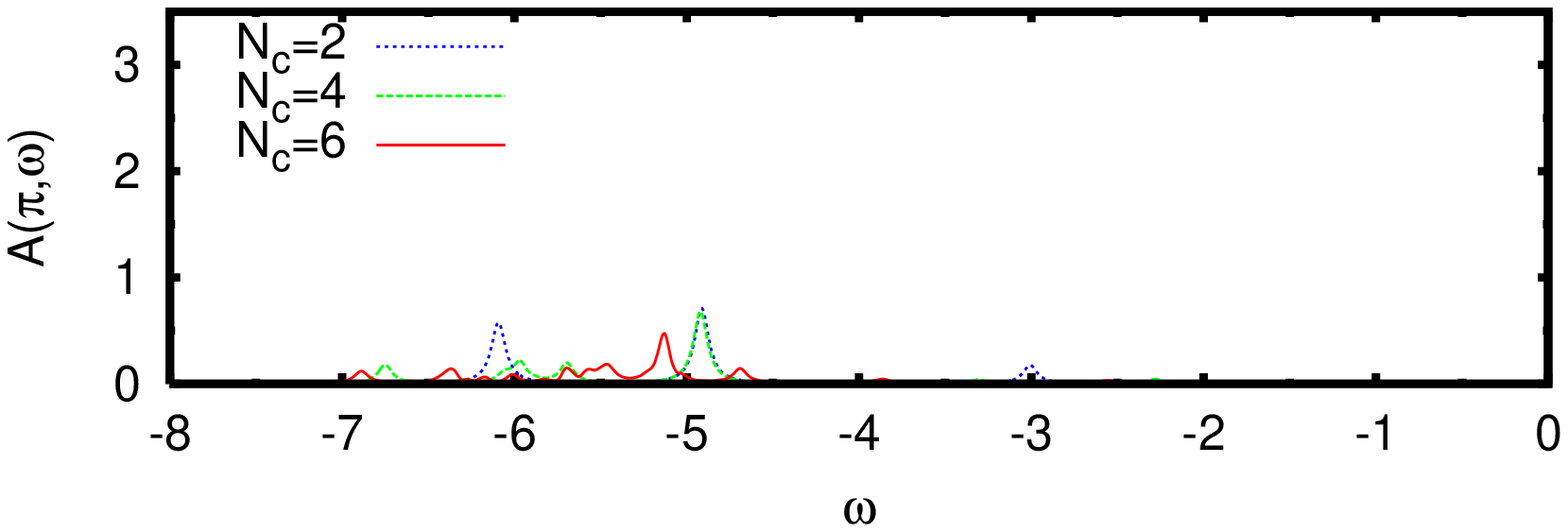}}
    \caption{\label{fig.spch}
Spectral weight $A(k,\omega)$ for $k=0,\pi/4,\pi/2,3\pi/4,\pi$ and $(N_c,N_b)=(2,8),~(4,8),~(6,6)$.
We used a broadening factor $\varepsilon=0.05$.
    }
\end{figure}

The plots of $A(k,\omega)$ at some $k$ points for $U=8$ as a function of $N_c$ in 
Fig.~{\ref{fig.spch}} help us understand the distribution of spectral weights 
more clearly.
The spectral weight at $k=0$ shows two prominent peaks, one around $\omega=-3$ 
and the other around $\omega=-4.5$, associated with spinon and holon branches,
respectively. 
Other small peaks are found to form a continuum between the two peaks as $N_c$
is increased.
At $k=\pi/4$ the system exhibits most spectral weight between the branches
$A$ and $B$ with low-intensity continuum tails between the branches $B$ and $C$.
The spinon and holon branches merge at $k=\pi/2$, leaving one prominent peak
around $\omega=-2$ with a long tail up to the lower boundary of spinon-holon
continuum.
The spectral weights at $k=3\pi/4$ and $\pi$ also consistently show the
continuum between the branches $B$ and $D$. 
We find that
the magnitudes of the spectral weights between the spinon and the holon 
branches are much larger than those in other regions.
Our analysis verifies clearly that the CDMFT with a cluster of only 6 sites
gives a quantitatively accurate description of the spectral properties in the 
one-dimensional half-filled Hubbard model.

\section{Summary} \label{sec.summary}

We have investigated the properties of the one-dimensional Hubbard model 
in the half-filled case within the CDMFT. 
The exact diagonalization method has been chosen as an impurity solver. 
The spectral gap and the energy density which have been computed via the CDMFT 
even with very small clusters
have been found to be in remarkable agreement with the exact values.
The spin-charge separation, the inherent nature of 
the one-dimensional interacting systems, 
has also been observed in the CDMFT study of spectral weights. 
Furthermore,
the CDMFT results for the spectral weights are in excellent agreement 
with Bethe ansatz dispersions and
their intensity are quite consistent with the existing 
experimental and theoretical results.
It is not yet clear whether such remarkable accuracy in spectral weights
can be achieved by the CDMFT with small clusters away from half filling,
which would be an interesting future project.

\ack
This work was supported by the Korea Research Foundation Grant funded
by the Korean Government (MOEHRD, Basic Research Promotion Fund)
(KRF-2007-314-C00075).
GSJ thanks the Korea Institute for Advanced Study,
where part of this work was accomplished, for its hospitality during his visit.


\section*{References}
\begin{thebibliography}{99}

 \bibitem{Georges1996} A. Georges, G. Kotliar, W. Krauth, and  M. J. Rozenberg,
Rev. Mod. Phys. {\bf 68}, 13 (1996).

 \bibitem{Maier2005} T. Maier, M. Jarrell, T. Pruschke, and  M. H. Hettler, 
Rev Mod. Phys. {\bf 77}, 1027 (2005).

 \bibitem{Kotliar2001} G. Kotliar, S. Y. Savrasov, G. P\'{a}lsson, and  G. Biroli, 
Phys. Rev. Lett. {\bf 87}, 186401 (2001).

 \bibitem{Bolech2003} C. J. Bolech, S. S. Kancharla, and  G. Kotliar, 
Phys. Rev.  B {\bf 67}, 075110 (2003).

 \bibitem{Hettler2000} M.H. Hettler, M. Mukherjee, M. Jarrell, and  
H.R. Krishnamurthy, Phys. Rev. B {\bf 61}, 12739 (2000).

 \bibitem{Potthoff2003} M. Potthoff, M. Aichhorn, C. Dahnken, Phys. Rev. Lett.
{\bf 91}, 206402 (2003).

 \bibitem{Rubtsov2008} A. N. Rubtsov, M. I. Katsnelson, and  A. I. Lichtenstein, 
Phys. Rev. B {\bf 77}, 033101 (2008).

 \bibitem{Toschi2007} A. Toschi, A. A. Katanin, and  K. Held, 
Phys. Rev. B {\bf 75}, 045118 (2007).

 \bibitem{Lieb1968} E.H. Lieb and F.Y. Wu, Phys. Rev. Lett. {\bf 21}, 
192 (1968).  

 \bibitem{Ovchinnikov1970} A.A. Ovchinnikov, Sov. Phys. JETP {\bf 30}, 
1160 (1970).

 \bibitem{Sorella1992} S. Sorella and A. Parola, J. Phys.: Condens. Matter 
{\bf 4}, 3589 (1992).

 \bibitem{Penc1996} K. Penc, K. Hallberg, F. Mila, and  H. Shiba, 
Phys. Rev. Lett. {\bf 77}, 1390 (1996).


 \bibitem{Benthien2004} H. Benthien, F. Gebhard, and  E. Jeckelmann, 
Phys. Rev.  Lett. {\bf 92}, 256401 (2004).

 \bibitem{Benthien2007} H. Benthien and E. Jeckelmann, Phys. Rev. B {\bf 75}, 20 5128 (2007).

 \bibitem{Senechal2000} D. S\'{e}n\'{e}chal, D. Perez, and  M. Pioro-Ladri\`ere, Phys. Rev. Lett. {\bf 84}, 522 (2000).

 \bibitem{Kim2006} B. J. Kim, H. Koh, E. Rotenberg, S.-J. Oh, H. Eisaki, N.
Motoyama, S. Uchida, T. Tohyama, S. Maekawa, Z.-X. Shen, and  C. Kim, Nature
Physics {\bf 2}, 397 (2006). 

\bibitem{Koitzsch2006} A. Koitzsch, S. V. Borisenko, J. Geck, V. B. Zabolotnyy,
M. Knupfer, J. Fink, P. Ribeiro, B. B\"uchner, and R. Follath,
Phys. Rev. B {\bf 73}, 201101(R) (2006).


\end{thebibliography}
\end{document}